\documentclass{aa}

\usepackage{graphicx}
\usepackage{txfonts}
\usepackage{subfigure}

\begin{document}

\title{Binary asteroid dissociation and accretion around white dwarfs}

   \author{Zeping Jin
          \inst{1}
          \orcid{0000-0001-8747-7407}
          \and
          Daohai Li\inst{1}\orcid{0000-0002-8683-1758}
          \and
          Zong-Hong Zhu\inst{1,2}\orcid{0000-0002-3567-6743}
          }

   \institute{Department of Astronomy, Beijing Normal University, Beijing 100875, China\\
              \email{202021160012@mail.bnu.edu.cn; lidaohai@bnu.edu.cn; zhuzh@bnu.edu.cn}
         \and
             School of Physics and Technology, Wuhan University, Wuhan 430072, China\\
             }

   \date{  Received 20 January 2023 /
   Accepted 31 March 2023
   }

\abstract{
About 25-50\% of white dwarfs (WDs) show metal lines in their spectra. Among the widely accepted explanations for this effect is that the these WDs are accreting asteroids that are perhaps flung onto the WDs by a planet via resonance, for instance. A number of theoretical works have looked into the accretion of asteroids onto WDs and obtained a fair agreement with the observed accretion rate. 
However, it is solely a very recent study (referenced in this work) that has taken  asteroid binarity into consideration, examining the scattering between an asteroid binary and planets and showing that a dissociation and ejection of the former might result and the effect on WD metal accretion is likely to be weak. Here, we investigate the close encounter between an asteroid binary and the central WD and consider how the binary's dissociation may affect the WD's accretion. We find that depending on the orbital and physical properties, the components may acquire orbits that are significantly different (even on the order of unity) from that of the parent binary. We assumed all the inner main belt asteroids are binaries and we studied their accretion onto the solar WD under the perturbation of the giant planets. We find that compared to the case without binaries, the components' accretion may be postponed (or put forward) by millions of years or more, as the objects may be taken out of (or driven deeper into) the resonance due to the sudden orbital change upon dissociation. However, the overall influence of binary dissociation on the accretion rate is not very significant.}

\maketitle

\section{Introduction} \label{sec:intro}
Spectroscopic observations have revealed metal lines in more than 1000 white dwarfs (WDs), with the estimated occurrence rate ranging between $25\%-50\%$ \citep[e.g.,][]{1980ApJ...242..195C,1982A&A...113L..13K,1983ApJ...272..660L,1986A&A...155..356Z,1997ASSL..214..293K,2003ApJ...596..477Z,2010ApJ...709..950K,2010MNRAS.404L..40V,2010ApJ...722..725Z,2012MNRAS.424..464F,2014MNRAS.444.1821F,2017ApJ...834....1M}. Since the effect of gravitational settling can stratify chemical elements in white dwarf atmospheres \citep{2006A&A...453.1051K} on timescales that are many orders of magnitude shorter than the white dwarf cooling time \citep{1986ApJS...61..177P,2006A&A...453.1051K}, these polluted white dwarfs must be continuously accreting metal-rich material in order to explain the observability of these metal lines. This phenomenon is referred to as the "metal pollution" of white dwarfs.

Several theoretical models have been proposed to explain metal pollution in this context. 
The currently favored mechanism is tidal disruption of metal-rich planetary material following a very close encounter with the central WD and subsequent accretion onto the host star \citep{2006Sci...314.1908G,2006ApJ...646..474K,2007ApJ...662..544V,2009ApJ...694..805F,2009ApJ...699.1473J,2010ApJ...714.1386F,2010ApJ...722.1078M,2017MNRAS.468..154B,2017MNRAS.468.1575B,2018MNRAS.474.4795X,2021MNRAS.508.5671L}.
Potential polluters could be asteroids \citep{2003ApJ...584L..91J,2006ApJ...653..613J,2009ApJ...699.1473J,2012ApJ...747..148D,2013MNRAS.435.2416V,2014MNRAS.439.3371W},
comets\citep{2017MNRAS.469.2750C}, 
or satellites \citep{2016MNRAS.457..217P,2017MNRAS.464.2557P}.

The planetary material could be flung towards the WD from planets \citep{2014MNRAS.439.2442F,2018MNRAS.476.3939M,2021MNRAS.506.1148V}. The Kozai-Lidov instabilities from stellar binaries {may} cause a perturbation {among} the planetary material\citep{2016MNRAS.462L..84H, 2017ApJ...834..116P}.
{If the magnetic field is taken into account}, Alfvén-wave drag is also an important mechanism \citep{2021ApJ...915...91Z}.

Binaries make up a significant fraction of Solar System small bodies. For example, nearly $16\%$ near-Earth asteroids larger than 200 meters may be binary systems\citep{2002Sci...296.1445M}; almost all of the planetesimals born near the Kuiper belt are binaries\citep{2017NatAs...1E..88F}. Yet, to our knowledge, previous works examining WD metal pollution by asteroid accretion \citep{2018MNRAS.480...57S, 2022ApJ...924...61L} have not taken asteroid binarity into consideration, {except in the very recent work by \citet{10.1093/mnras/stad382}. {In that work, the authors examined the binary asteroid scattering with planets around a WD and showed that the binary may be dissociated or even ejected from the system; neither process has a significant effect on the accretion rate of WDs. In this work, we instead look at the scattering between an asteroid binary not with a planet but with the WD itself. The rationale is outlined below.} }

In the accretion mechanism discussed above, an asteroid first enters the Roche lobe of the WD, gets tidally disrupted (physically deformed and stretched into a ring of constituent material), and is then accreted. For a binary of asteroids, en route to the tidal disruption of an individual component, the binary would have been tidally dissociated first and the two constituents would become gravitationally unbound (but {both are likely to remain} physically intact) and move on their respective orbits. {We are interested in whether this process affects WD metal pollution.}

For instance, secular resonances may excite the orbital eccentricity of an asteroid and send it to the central WD to be accreted \citep{2018MNRAS.480...57S}. If the asteroid is a single object, it will be forced onto the WD (or towards tidal disruption) {uninterruptedly}. However, should it be a binary, as we go on to show in this work, the two components would be dissociated well before the tidal disruption; in addition, because the two objects have different velocities and positions with respect to the WD upon the dissociation, they will acquire orbits that are different from each other and from that of the parent binary. Therefore, if the change in the component's trajectory is significant after the dissociation, it could be no longer affected by the secular resonance {and this may or may not quench the accretion.}

The paper is organized as follows. In Section \ref{sec:ms2}, we describe the binary dissociation with both theoretical derivation{s} and numerical simulation{s}. In Section \ref{sec:ms3}, we perform a series of simulations to test {how this affects the} WD accretion. 
Section \ref{sec:Con} presents our conclusions and discussion.

\section{Binary dissociation} \label{sec:ms2}
Here, we consider a three-body problem: the dissociation of an asteroid binary around a star. We derive the change in the component's orbit after the dissociation using basic principles and then verify this based on numerical simulations.

\subsection{Analytical derivations}

Supposing a binary asteroid is approaching the central host star of mass M on an orbit of semimajor axis, $a_{tot}$, and eccentricity, $e_{tot}$, with respect to the central object. The two components' (referred as Ast1 and Ast2) masses are $m_1, m_2$, their relative orbit being $a_{rel}$. {In this work, the subscript "tot" is used to describe the quantities related to the motion of the barycenter of the binary around the star and "rel" the relative motion of one binary component with respect to the other.} The two objects’ relative velocity is then
\begin{equation}
v_{rel} \sim \sqrt{\frac{G\left(m_1+m_2\right)}{a_{rel}}},
\end{equation}
where $G$ is the gravitational constant. Supposing the binary asteroid dissociates at a heliocentric distance of $r_{d}$, we further assume
that $r_d<<a_{tot}$ and $r_d>>a_{rel}$ {(justified later in this paper)}. 
Upon dissociation, the binary asteroid's energy fulfills the following:
\begin{equation}
\frac{v_{tot}^2}{2}-\frac{G M}{r_d}=E=\frac{-G M}{2 a_{tot}}
,\end{equation}
where $v_{tot}$ is the velocity of binary at the instant of dissociation. The velocity of each binary component is $v_{tot} \pm v_{rel}$, at a distance to the host of $r_d \pm a_{rel}$. Then the specific energy of Ast1 at the moment of dissociation is:
\begin{equation}
\frac{\left(v_{tot} \pm v_{rel}\right)^2}{2}-\frac{G M}{r_d \pm a_{rel}}=-\frac{G M}{2 a_1}
.\end{equation}
{Subtracting the two sides of the above two equations respectively and making an expansion in $a_{rel}/r_d$, we obtain}
\begin{equation}
\pm v_{tot} \sqrt{\frac{G\left(m_1+m_2\right)}{a_{rel}}} \pm \frac{G M}{r_d^2} a_{rel}=-\frac{G M}{2}\left(1 / a_{tot}-1 / a_1\right).
\end{equation}
Furthermore, $v_{tot} \sim \sqrt{\frac{2 G M}{r_d}}$ , whereby the binary {total} orbit is very eccentric {and we take it to be parabolic}, so
\begin{equation}
\pm \sqrt{\frac{2 M}{r_d}} \sqrt{\frac{\left(m_1+m_2\right)}{a_{rel}}} \pm \frac{M}{r_d^2} a_{rel}\sim-\frac{M}{2}\left(1 / a_{tot}-1 / a_1\right)
.\end{equation}
{The value of $r_d$ can be estimated following \citet{2006Natur.441..192A}}
\begin{equation}
r_d \sim a_{rel}\left(\frac{3 M}{m_1+m_2}\right)^{1 / 3}.
\end{equation}
{It is clear from the above expression that $r_d\gg a_{rel}$. Then we have}
\begin{equation}
\pm\left(\frac{m_1+m_2}{M}\right)^{2 / 3} \frac{1}{a_{rel}} \sim-\frac{1}{2}\left(1 / a_{tot}-1 / a_1\right).
\label{eq m1}
\end{equation}
This means that the change in the binary's semi-major axis of each object with respect to that of the parent binary depends on the total mass of the binary positively and the semi-major axis of relative orbit negatively. The binary {total} orbit also plays a role.

In the above derivation, we have omitted the mass ratio of the binary components {$\mu=m_2/m_1$}. When $\mu$ is taken into consideration, Ast1 and Ast2 may obtain different orbits after the binary dissociation. For instance, the expression for Ast1 is
\begin{equation}
\pm\left(\frac{\mu}{1+\mu}\right)\left(\frac{m_1+m_2}{M}\right)^{2 / 3} \frac{1}{a_{rel}} \sim-\frac{1}{2}\left(1 / a_{tot}-1 / a_1\right).
\label{eq mu a1}
\end{equation}
Details {of the derivation} can be seen in Appendix \ref{ap1}.

\subsection{Numerical simulations}
Now, we perform a series of simulations for binary asteroid dissociation. All of the simulations in this work are carried out with MERCURY\citep{1999MNRAS.304..793C}. MERCURY is a general-purpose software package for N-body integrations. It is designed to track the orbital evolution of objects moving in the gravitational field of a large central body. We used the BS integrator with a tolerance of $10^{-12}$.

Our numerical model includes a central WD of $0.54$ solar masses ($M_{\odot}$) and a binary asteroid. We initiated the binary such that its center of mass is either 10 au from the WD or at the apastron, whichever is {smaller}. The periastron and apastron distances of the binary barycenter, semimajor axis, and eccentricity of the binary's relative orbit, the radius of Ast1, and the mass ratio $\mu$ are varied on a grid (as shown in Table \ref{table1}). 
{Each parameter combination} is run 100 times with different random numbers: 
{the remaining four orbital elements are initialized randomly and the norm of the binary relative orbit is isotropic.}
So, we have a total of $729600=1209600-480000$ runs. Here, $1209600=7 \times 8 \times 4 \times 3 \times 6 \times 3 \times 100,$ but 480000 of those are not self-consistent: for example, the sum of the radii of two components is larger than their relative orbital semi-major axis.

\begin{table*}[!t]
    \caption{The parameter set for grid simulations}
    \centering
    \begin{tabular}{lc}
    \hline\hline
        Parameter & {values}  \\\hline
    {Periastron} distance of total orbit(AU) & $0.006,0.01,0.014,0.016,0.02,0.024,0.028,0.032$\\
    {Apastron} distance of total orbit(AU) & $5,10,40,200,1000,5000,30000$\\
    Radii of Ast1(km) & $1,10,100,1000$ \\
    $\mu$ & $0.01,0.1,0.99$ \\
    semi-major axis of relative orbit(km) & $2,20,200,2000,20000,200000$ \\
    Eccentricity of relative orbit & $0.0,0.3,0.9$\\\hline
    \end{tabular}
    \tablefoot{Ast1 and Ast2 are the two members of a pair of binary {and} $\mu$ is the mass ratio of Ast2 to Ast1 {(Ast2/Ast1)}{. }
    For the total orbit, the mean anomaly {of the total orbit} is set such the binary is at a distance of $10AU$ from the central host or at the apastron if its {total} orbit is small {and} the remaining 3 {total orbital} elements {(inclination, argument of periastron and longitude of ascending node)}
    will be set {to} $0$. For the relative orbit, the remaining 4 {orbital elements} will be randomly chosen{.} For each parameter set, 100 runs are performed, implying a total of $1209600$ parameter-sets.}
    \label{table1}
\end{table*}

Each run lasts for an orbital period of the binary's total orbit so upon the termination of the simulation, the small bodies are sufficiently far from the WD. Then we calculate the orbital elements of each component with respect to each other and the WD. If the two components are bound, we additionally calculate the heliocentric orbit of the barycenter. The result shows that 563373 sets of 729600 binaries have dissociated {during the encounter with the WD}. {These details are further examined in Sect. \ref{sec:k}.}

\begin{figure}[h]
    \centering
    \subfigure[]{
    \includegraphics[scale=0.5]{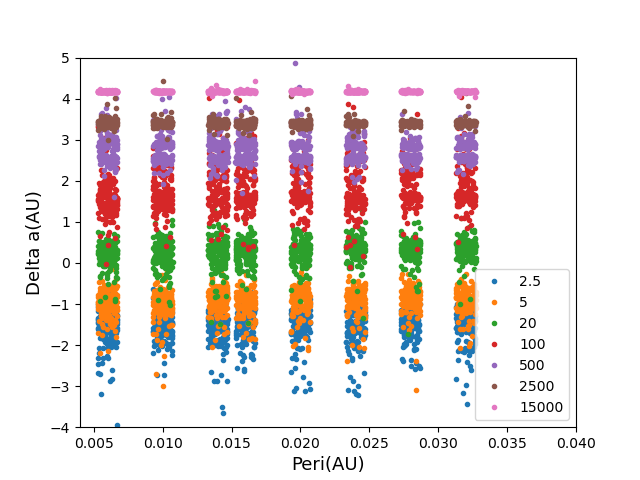}
    }
    \subfigure[]{
    \includegraphics[scale=0.5]{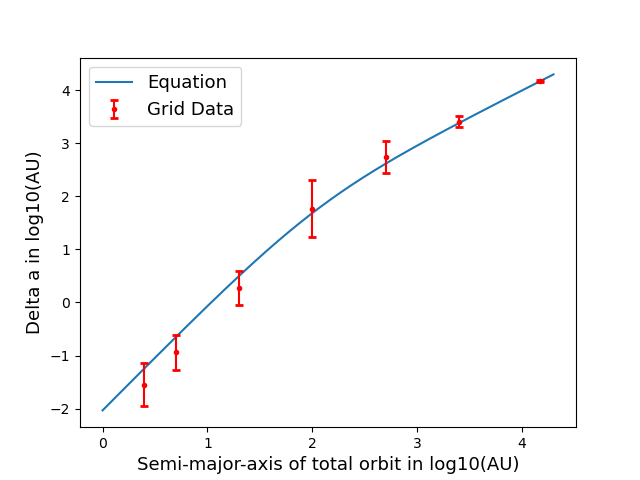} 
    }
    \caption{Dependence of $\Delta a$(For Ast1) on the binary asteroid's total orbit. Panel (a) shows the result of our grid simulation, where the size of Ast1 is 100km, $\mu=0.99$ and the semi-major axis of the relative orbit is 2000km. The x-axis is the {periastron} distance of the binary's total orbit, and the y-axis is the change in semi-major axis (Ast1) which is presented in log scale. The different colors mean the different semi-major axis of the binaries' total orbit. The dots have been shifted horizontally by a small but random amount for better visibility.
    Panel (b) shows the relationship between $a_{tot}-a_1$ and $a_{tot}$. The blue line comes from Equation \eqref{eq m1}. The red dots are the mean value of our grid simulation and the error here is the standard deviation. }
    \label{fig:grid}
\end{figure}

{First,} we concentrate on the change in the orbits of $\Delta a${, the absolute value of the difference between the semi-major axis of Ast1 or Ast2 and that of the parent binary's total orbit}. 
Figure \ref{fig:grid} displays $\Delta a$ as a function of the {periastron} distance of the barycenter of the binary before the dissociation. Only the result from the radius of Ast1=100 km, $\mu=0.99$ runs are shown for clarity. The different colors are for different binary semimajor axes. {From Figure \ref{fig:grid}(a), we can see that $\Delta a$ does not change very much with the periastron distance of the total orbit and the semi-major axis does affect $\Delta a$ significantly. }

{Figure \ref{fig:grid}(b) displays the} relationship between {the} total semi-major axis and $\Delta a$, {where the simulations agree well with the analytical derivations Equation \eqref{eq m1}}. Additionally, the result also {shows} that $\Delta a$ can {acquire} a quite large value if the total semi-major axis is large.

\begin{figure}[h]
    \centering
    \subfigure[]{
    \includegraphics[scale=0.52]{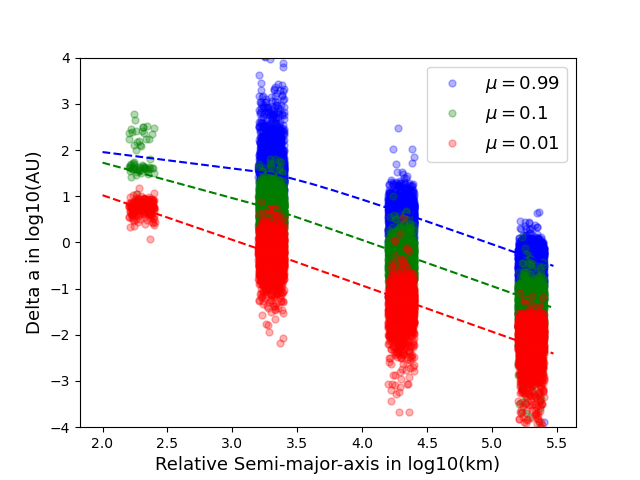} 
    }
    \subfigure[]{
    \includegraphics[scale=0.52]{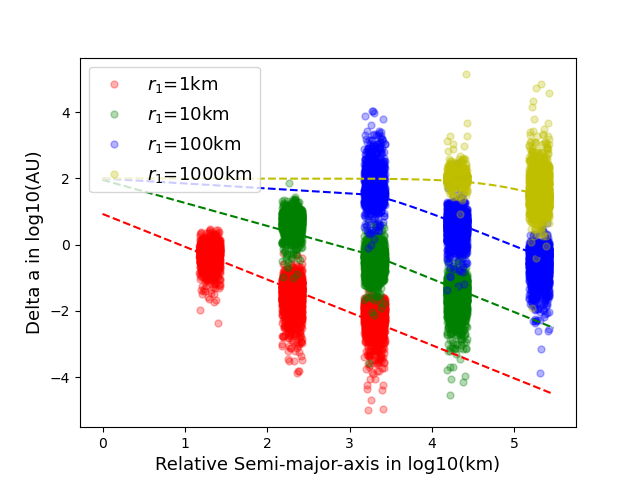}
    }
    \caption{Dependence of $\Delta a$(For Ast1) on the binary asteroid's relative orbit. The x-axis is the semi-major axis of binary relative orbit the y-axis is the change in the semi-major axis, both in log scale. Dots are from the grid simulations. To make the plot clearer, the dots have been shifted horizontally by a small but random amount (otherwise there are only four values for the x-axis: 200km, 2000km, 20000km and 200000km). For {panel (a)}, Different colors are used for different $\mu$ (red 0.01, green 0.1, blue 0.99). {The radii of Ast1$r_1$} is 100km. The three dashed lines are the predictions of Equation \eqref{eq mu a1}.
    For {panel (b), different} colors are used for different $r_1$(red 1km, green 10km, blue 100km, yellow 1000km). {The mass ratio} $\mu$ is 0.99. The four dashed lines are the predictions of Equation \eqref{eq m1}.
    }
    \label{fig:grid2}
\end{figure}

The dependence of $\Delta a$ on the relative semi-major axis, the total mass of the binary, and $\mu$ is shown in Figure \ref{fig:grid2}. 
In panel (b), the simulation result agrees with Equation \eqref{eq m1}: the heavier the total mass or the tighter the binary, the larger the change $\Delta a$. 
The result in (a) is also compatible with Equation \eqref{eq mu a1}: the larger the mass ratio ($\mu$) or the tighter the binary, the larger $\Delta a$ for Ast1.

During our simulation, ejections are recorded in 39727 runs. In all of these, the smaller component Ast2 is ejected and Ast1 remains bound to the WD. A comparison between $\Delta a$ for Ast1 \eqref{eq mu a1 app} and that for Ast2 \eqref{eq mu a2 app} shows that when the mass of Ast2 is small ($\mu$ is small), the dissociation will cause a negligible change in the orbit of Ast1, but the alternation in Ast2's orbit is always close to the case where the two have comparable masses, the reason why only the ejection of Ast2 is possible. We note that ejection caused by scattering with planets might be more frequent \citep{10.1093/mnras/stad382}.

\subsection{Dissociation distance} \label{sec:k}
In the derivations in Section 2.1, we assume that the binary dissociates at a distance of $r_{\mathrm{d}}=a_{\mathrm{rel}}\left(\frac{3 M}{m_1+m_2}\right)^{1 / 3}$ from the central host, as suggested by \citet{2006Natur.441..192A}. This is perhaps better understood as a scale factor rather than the exact dissociation distance.  In order to {examine} this effect, we introduced a factor $k$
\begin{equation}
    s_{\mathrm{d}}(k)=k{\cdot}a_{\mathrm{rel}}\left(\frac{3 M}{m_1+m_2}\right)^{1 / 3}.
\label{eq_k}
\end{equation}
In the following, we intend to put a constraint on k using numerical simulations.

Hundreds of Solar System binary objects have been measured with a fair level of accuracy, as compiled by W. R. Johnston\footnote{\url{https://sbn.psi.edu/pds/resource/doi/binmp_3.0.html}}\citep{2019pdss.data....4J}, using observations by \citet{2008Icar..193...74D,2005AJ....129.1117E,2017pdss.data....3H,2002aste.book..289M,2006IAUS..229..301N,2008ssbn.book..345N,2006Icar..181...63P,2007Icar..190..250P,2006PDSS...50.....R,2009EM&P..105..193W}. 
From those, we {chose pairs that meet the criterion:} the radius of Ast1 is in the range (50km,1000km) and the semi-major axis of their relative orbit is within (500km,2000km). {A total of 42 asteroid binaries were thus chosen.}

In our simulations, we used the physical parameters and their relative orbits. As for the heliocentric orbit of the binary center of mass, the {apastron} distance is $5AU$ and the {periastron} distance is set {by using} $k$. Twenty values for $k$ evenly distributed between $0.01$ and $2.0$ were tested. The angular orbital elements are drawn randomly. For each $k$, we carried out 100 runs with different random numbers.

\begin{figure}[h]
    \centering
    \includegraphics[scale=0.55]{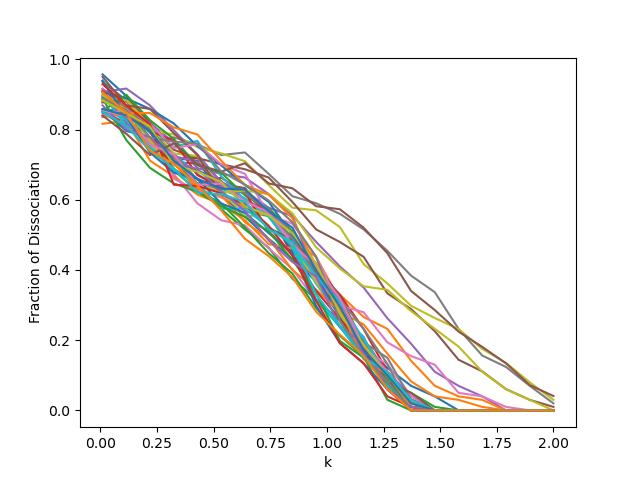}
    \caption{Fraction of systems that dissociate as a function of k, as defined in Equation \eqref{eq_k}. Different colors represent the 42 sets of asteroid parameters. 
    }
    \label{fig1}
\end{figure}

In Figure \ref{fig1}, we show the fraction of dissociated binaries for each $k$ and for each binary pair chosen as above. The figure {shows that the dissociation rate is decreasing as $k$ increases}. We note that at $k=1$, the dissociation fraction is only about $0.3$, which is significantly different from \citet{2006Natur.441..192A}($>0.95$). This could result from the different setups for the inclinations in the two works. 
{For example, the inclination has been shown to greatly affect the outcome of few-body scatterings \citep{2005A&A...437..967P}; and  while \citet{2006Natur.441..192A} looked into coplanar encounters only here we are examing isotropic encounters.} 
Also, at $k=0.75$, about half of the binaries dissociate.

\begin{figure}[h]
    \centering
    \includegraphics[scale=0.55]{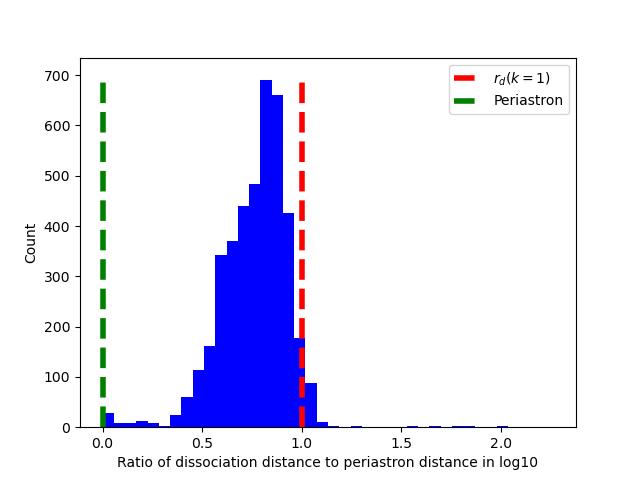}
    \caption{
    {Distribution of the ratio of disruption distance to periastron distance. 
    The x-axis is logarithmic of the dissociation distance over periastron distance.
    Red dashed line represents the distance of $s_d(k=1)$, and the green dashed line is the periastron distance. 
    In this simulation, we set the periastron distance as $0.1s_d(k=1)$.}
    }
    \label{fig dist}
\end{figure}

It is hard to locate the exact location of dissociation, as the binary switches back and forth between unbound and bound relative motion a few times before eventually being torn apart.
To gain an understanding of the dissociation distance, we perform a new simulation during which we record the last moment when binaries' relative orbital eccentricity is smaller than 1 and take this as the dissociation time and the heliocentric distance of the binary barycenter as the dissociation distance. In this simulation, we still use the 42 binaries as before but now the relative inclination is set to 0 to minimize its effect on the dissociation rate. For the total orbit, the periastron distance of the total orbit is now $0.1s_d(k=1)$.
Figure \ref{fig dist} shows the distribution of the dissociation distance.
The x-axis is the ratio of dissociation distance to the periastron distance of the binaries' total orbit(in log scale). The red vertical dashed line marks the heliocentric distance $s_d(k=1)$ and the green one shows the periastron distance.
The plot shows that most binaries do not dissociate at the periastron (x=0) but at an x lower than 1. This implies the majority of the binaries get dissociated between the periastron and $s_d(k=1)$. 

We have shown that when a binary asteroid travels close to the central host star, it may be dissociated and the two components may acquire orbits that are significantly from that of the parent binary. In the next section, we assess how this {affects} the WD accretion rate.

\section{Effect on the WD accretion rate} \label{sec:ms3}
In order to study the fate of the dissociated binaries, we need a ``{template}'' planetary system that drives the binary close to the WD in the first place. Here, we take the Solar System as an example. \citet{2022ApJ...924...61L} modeled the evolution of the Solar System (giant planets + small bodies) during the giant branch and the WD phase. Here we are interested only in the WD phase of the Sun's evolution. During the giant branch, the solar radius expands and the Sun loses its mass \citep{1977A&A....57..395R,2015MNRAS.448..502M,2014ApJ...790...22R,2016ApJ...822...73R}. This causes the planet and the small body orbits to adiabatically expand \citep{2018MNRAS.480...57S}, conserving their respective angular momentum \citep{1998Icar..134..303D}. We take the planets' and the small bodies' state vectors from \citet{2022ApJ...924...61L} at the beginning of the WD stage. Thus our model contains the Sun, Jupiter, Saturn, Uranus, Neptune, and small bodies of the Solar System. We restrict ourselves to only small bodies in the inner main belt (inner to 2.5 AU), where resonances play a major role in driving the objects to the central host \citep{2022ApJ...924...61L}.

Our simulations are divided into three stages. First, we follow the evolution of the binary center of mass and the relative motion of the two components is ignored; thus the binary is actually treated as a single {object}. Here, we refer to this step as the initial orbit simulation.
    Second, we model the binary's dissociation. The two components are inserted into the simulation as their barycenter approaches the central WD. We call this step the dissociation simulation.
    Third, we study the long-term evolution of the dissociated binaries and put an emphasis on {whether} and when a component collides with the central WD. Here, we call this step {the accretion} simulation.

\subsection{Initial orbit simulation}

We consider all main belt asteroids with the following properties: (1) the absolute magnitude is brighter than $12.8$ so the number of objects is within the reach of our computational capacity and (2) the semimajor axis is within the range $(2,2.5)$ au so that the objects have a fair chance of being affected by the $\nu_6$ secular resonance and thrown close to the WD. {Those objects have been picked from the minor planet center\footnote{\url{https://minorplanetcenter.net}} by \citet{2022ApJ...924...61L} and run through the Sun's giant branch (assuming a linear mass loss within 20 Myr) after which 605 survive. These heliocentric orbits of those 605 objects are the ones used in our simulation.} Here, we are interested only in the WD phase evolution, so we take the state vectors of our chosen objects from \citet{2022ApJ...924...61L}. Our simulations are run for $5\times10^8$ yr.

We emphasize that in this stage of simulation, we only take the asteroids' orbital parameters but not their physical properties. Suppose an asteroid binary is under investigation. If the binary barycenter is sufficiently far from the WD (and planets), the two components revolve around the center of mass much like that prescribed by a two-body problem. On the other hand, the motion of the barycenter is controlled by the Sun and perturbed by the planets in the system. In other words, the relative motion and the barycenter motion can be separated and tackled one by one. In this stage of the simulation, we are interested in how the barycenter moves. Thus, all asteroids are treated as massless small bodies. However, when an asteroid's heliocentric distance becomes very small, the binary may be dissociated by the WD, so we record the state vectors of that object and all planets when the object's distance to the WD is $<$ 0.2 AU. Yet at this stage of simulation, the object is kept in the simulation until it is accreted by the WD or is ejected. The state vectors of the objects are  then used in the next stage of our simulation.

\begin{figure}[]
    \centering
    \includegraphics[scale=0.55]{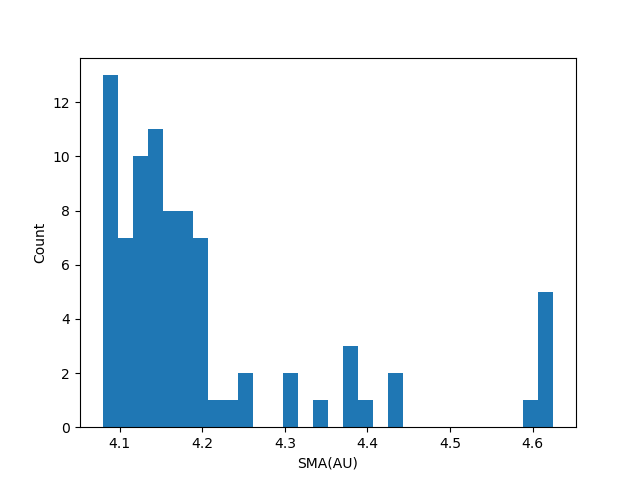}
    \caption{Distribution of semi-major axis(after orbital expansion) of asteroids whose {periastron} distance has ever been lower than $0.2AU$ {in the initial orbit simulation}.} 
    \label{fig3}
\end{figure}

From these simulations, we obtained the state vectors of all small bodies (and the planets) when their {periastron} distance dips under 0.2 AU. Among the 605 asteroid objects tracked in the simulation, 83 have approached the WD closer than 0.2 AU and are recorded. Figure \ref{fig3} shows the distribution of the semimajor axes of these objects. Most of the objects are within $4.1AU$ and $4.2AU$, close to the location of the $\nu_6$ resonance after the orbital expansion\citep{2018MNRAS.480...57S,2021MNRAS.504.3375S}. {No ejections are recorded in this stage of the simulation as even after the eccentricity is pumped up, the asteroids' orbits are still decoupled from those of the planets so a close encounter with them is unlikely.}

\subsection{Dissociation simulation}

From the previous simulation, we obtained the orbit of the small body as it approaches the WD. Now we are ready to investigate their dissociation and need to take the relative motion into account. 
As detailed above, 42 observed binaries were selected, and their physical sizes and {relative} orbital semimajor {axis} and eccentricity are used to set their relative motion. 
The barycenter follows the trajectory we obtained from the previous integration.
{To put it simply, the total orbit comes from the previous simulation, while the relative orbit, physical size, and mass come from the observations.}  {Specifically, we used the relative orbital period and separation of the 42 binaries to calculate the total mass and then use their size to get the individual mass, all available in the observational data set.}

However, simulating all {of the binary total orbits recorded in the previous simulation} is beyond our computing power. So, 16 of them were randomly chosen from {the 83 pairs whereby the pericenter distance has fallen below 0.2 au at some point during the} initial orbit simulation, all with semi-major axes in {the} range of (4.1,4.2) AU. 
Therefore, we obtained 42*16 parameter setups of binary dissociation simulation. For each setup, we further randomized the phase angles of the binary relative 100 times, implying a total of $16*42*100=67200$ runs.

{Notably}, the dissociation distance depends on the asteroids' properties (Eq \eqref{eq_k}), and the 42 binaries thus have different $r_d$. Section 2.3 shows that the chance of dissociation is 0.5 when $k=0.75$. So we let the barycenter orbit be that when {the} {periastron} distance becomes smaller than $s_d(k=0.75)$ for the first time {in the initial orbit simulation}.

At this stage, the two components are treated as big bodies so their mutual gravity is {taken into account}. The simulation is run for about a period of the binary total orbit {so the dissolution can be resolved. Altogether, 36732 of 67200 binaries are dissociated. {Below we analyze the statistics of those dissociated.}

Depending on the mass ratio of two objects of the binary, we can divide them into two classes. The first class is with components of almost equal sizes; here, we require that $\mu>0.1$. The second class is the satellite system where $\mu<0.1$. These two classes may have different formation scenarios \citep{2021PSJ.....2...27N}.

As we show above, after a binary asteroid's dissociation, the two components may acquire semimajor axes that are different from that of the parent binary. This may take the object out of resonance and quench further eccentricity excitation or, otherwise, drive it deeper into the resonance. The change in the semi-major axis after the dissociation is presented in Figure \ref{fig2}.

Figure \ref{fig2} shows that for most of Ast1 and Ast2, the semi-major axis will change by more than $0.01AU$.
{The lower panel clearly shows two peaks which, as the lower two rows display, represent the dissociation of similar mass binaries ($\mu>0.1$, blue) and of dissimilar mass binaries ($\mu<0.1$, orange).}
In the latter case, the change in the semimajor of Ast2 is larger (as predicted by Equation \eqref{eq mu a1}) and can often be even larger than $0.1AU,$ {when $\mu<0.1$}, comparable to the width of the $\nu_6$ resonance \citep{2021MNRAS.504.3375S}. The distribution of Delta a for Ast1 also shows two peaks, but the two are not well separated  by $\mu=0.1$.

\begin{figure}[]
    \centering
    \includegraphics[scale=0.6]{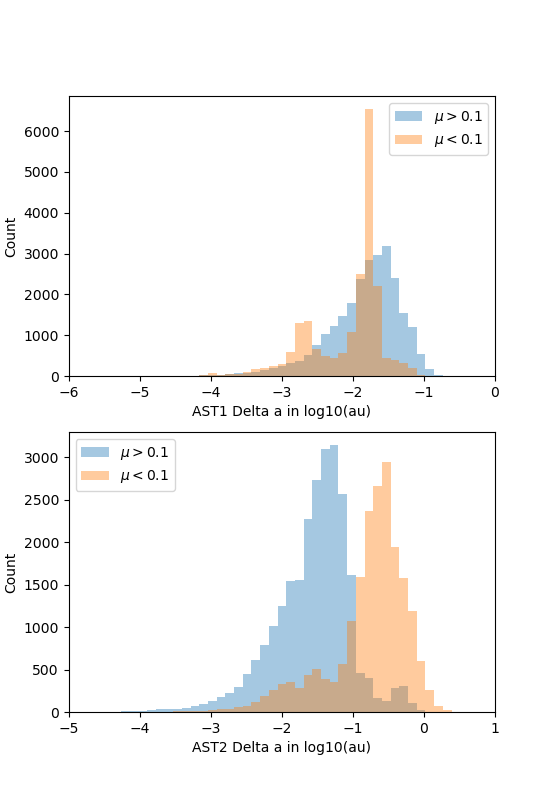}
    \caption{Distribution of the changes of the semi-major axis after the dissociation. {Upper panel: Distribution for Ast1 and the lower panel is for Ast2. Two ranges of $\mu$ are represented by different colors, blue for $\mu>0.1$ and orange for $\mu<0.1$.}
    The difference is {the logarithmic of the absolute value of the change}. ($log_{10}(|a_{\mathrm{after}}-a_{\mathrm{before}}|)$).} 
    \label{fig2}
\end{figure}

\subsection{{Accretion} simulation}
At this stage, we took both the components from the 36732 dissociated binaries, treating them as {independent} small bodies and further integrate their evolution under the effect of the planets for $2\times 10^{8}$ yr. In the simulation, we recorded the time at which the components Ast1 and Ast2 collide with the central WD, referring to it as the {accretion} time. 

In order to reveal the effect of binary dissociation on the {accretion} time, we ran an additional simulation, where the first stage (the initial orbit simulations) is extended also for $2\times 10^{8}$ yr and the {accretion} time recorded. Thus, the {accretion} times with and without binary dissociation could be compared.

\begin{figure*}[]
    \centering
    \includegraphics[scale=0.75]{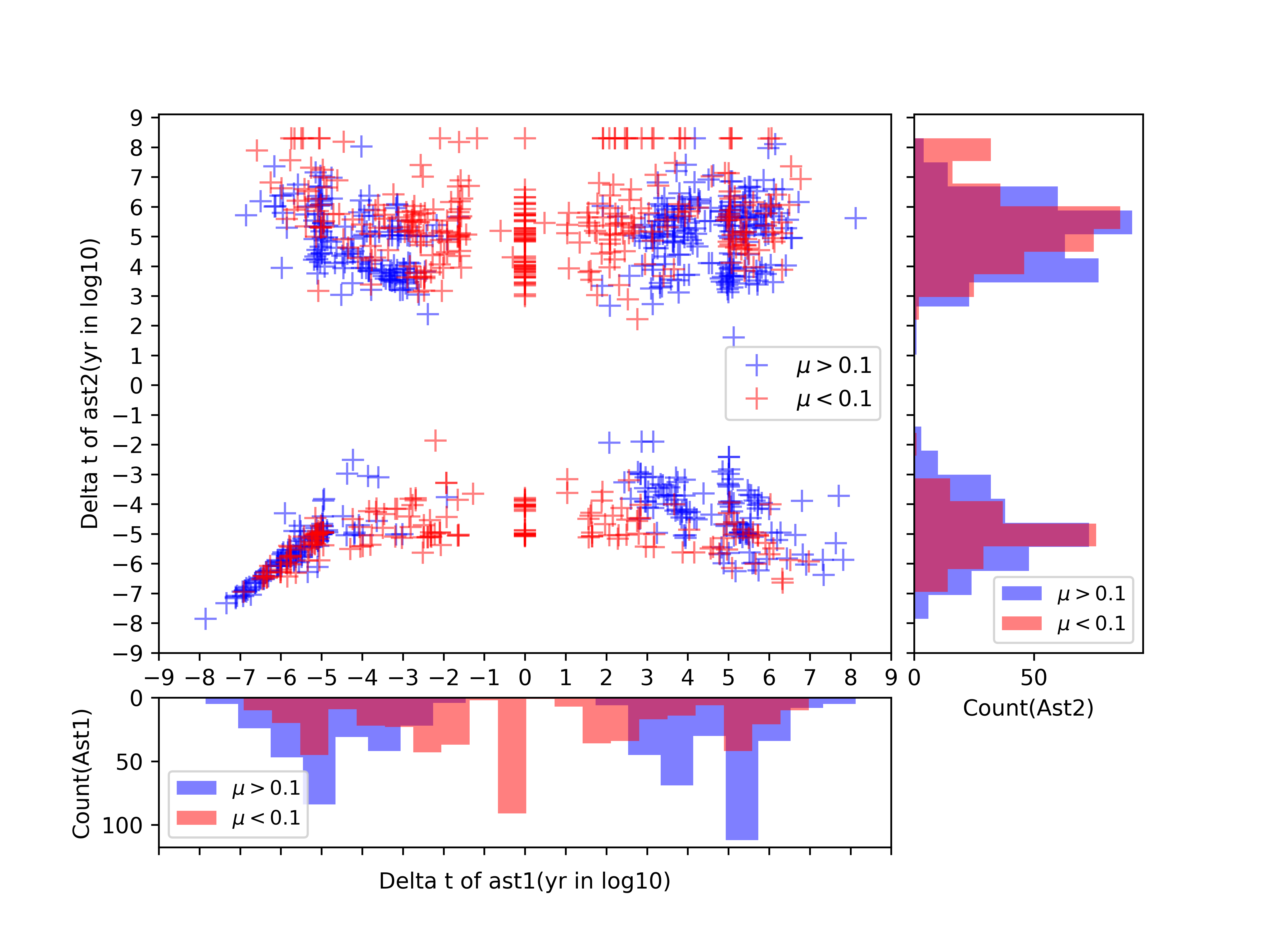}
    \caption{Change in the {accretion} time for Ast1 and Ast2. Center shows the major plot {where} the x-axis is the change in Ast1 and the y-axis is the change in Ast2. The axes are in log scale. All changes in the {accretion} time are larger than a year and a negative value here means that the {accretion} time is earlier than {without binary dissociation}. It can be explained by this equation:
    $x=\frac{\Delta t_{1}}{\sqrt{\Delta t_{1}^{2}}} \cdot \log (|\Delta t_1|)$ and $y=\frac{\Delta t_{2}}{\sqrt{\Delta t_{2}^{2}}} \cdot \log (|\Delta t_2|)$. Besides the major plot, the two histograms are the distribution of {the change} for Ast1(bottom) and Ast2(right), two colors representing different $\mu$. Their scales are the same as the main plot.}
    \label{fig5}
\end{figure*}

Figure \ref{fig5} shows the change in the difference in the {accretion} time with and without binary dissociation $\Delta t${ (difference between the accretion time of Ast1 or Ast2 and the original accretion time with no dissociation), the positive value means the accretion is delayed and the negative value means the accretion is {put forward}.}
In the large panel, the horizontal axis is $\Delta t$ of Ast1 and the vertical axis is that of Ast2. The blue and red colors are for $\mu>0.1$ and $\mu<0.1,$ respectively. The histograms besides the main plot are the distribution of $\Delta t$ for Ast1 and Ast2, respectively. {About} half of Ast1 and {nearly 2/3} Ast2 have their {accretion} time delayed, suggest{ing} that binary dissociation may have an effect on white dwarf metal pollution. The distribution of $\Delta t$ {for both} Ast1 and Ast2 shows two peaks at $\pm 10^5-10^6 yrs$. Additionally, for Ast1 and $\mu<0.1$, there is another peak at 0{where the orbit of Ast1 barely changes after the dissociation}.

The change in the {accretion} time should be related to the change in the semi-major axis, so we plot this relation in Figure \ref{fig6}, {which, again, uses} two different colors to highlight the effect of $\mu$. It is obvious when $\mu<0.1$ the change in Ast1 is smaller than that of Ast2. In this case, Ast2 is much smaller than Ast1 and may acquire a larger orbital change and this may prolong the {accretion} time further{, as can be expected from} Equation \eqref{eq mu a1 app} and Equation \eqref{eq mu a2 app}.
For Ast1, the result is obvious: when Ast1 obtains an orbit far from the resonance location ($4.1AU-4.2AU$), the {accretion} time is prolonged {or put forward appreciably potentially by} millions of years. The same can also be seen {for Ast2}. 
There exist some extreme cases where the accretion of Ast2 is postponed beyond our simulation of $2\times 10^{8}$ yr. This happens for $\mu<0.1$ and where $a_2>4.8$ au. When this occurs, we simply let $\Delta t=2\times 10^{8}$ yr so those points show up in the top right part of the bottom panel.

In order to compare our results with the observations, we calculate the accretion rate via:
\begin{equation}
    \dot{M}_{a c c 0}(t) =S \times (M_{1}(t) + M_{2}(t)) / T
,\end{equation}
{and}
\begin{equation}
    \dot{M}_{a c c 1}(t) =S \times (M_{1}(t) / T + M_{2}(t) / T), 
\end{equation}
{where $\dot{M}_{a c c 0}$ is the accretion rate without dissociation and  $\dot{M}_{a c c 1}$ is the accretion rate with dissociation, and both are functions of time $t$. Here, $T=3 \times 10^6$ yr is the length of time interval $(t-T,t)$ within which $M_{1}(t)$/$M_{2}(t)$ is the total mass of Ast1/Ast2 that is accreted by the WD. And $S$ is a scaling factor}
\begin{equation}
    S = \frac{M_{belt}}{M_{sample}}
,\end{equation}
which we use to re-scale the accretion rate of our simulation so the result corresponds to the real asteroid belt. Also, $M_{belt}$ is the total mass of asteroid belt, {and} we take $M_{belt}=18 \times 10^{-10} M_{\odot}$ here \citep{2002Icar..158...98K}. $M_{sample}$ is the total mass of {all} the asteroids in our simulations. $M_{sample}=605 \times M_{42}\times 100$ {and} $M_{42}$ is the total mass of the 42 real binaries we chose.

Figure \ref{fig:ac} shows the accretion rate. {The starting point  is the beginning of our Initial orbit simulation, so the time here also represents the cooling time.}. {T}he accretion without dissociation mostly occurs before $3 \times 10^7$ years, but the accretion with dissociation can occur until $10^8$ years, although at a much lower rate.
Overall, the effect of dissociation on accretion is rather insignificant.  
A closer examination seems to suggest that binary asteroids' dissociation can decrease the accretion rate somewhat, at cooling ages younger than $10^7$ yr. At a later time, the binary dissociation seems to boost the accretion rate appreciably, though the rate is anyway much smaller than at the early time.

\begin{figure}[]
    \centering
    \includegraphics[scale=0.57]{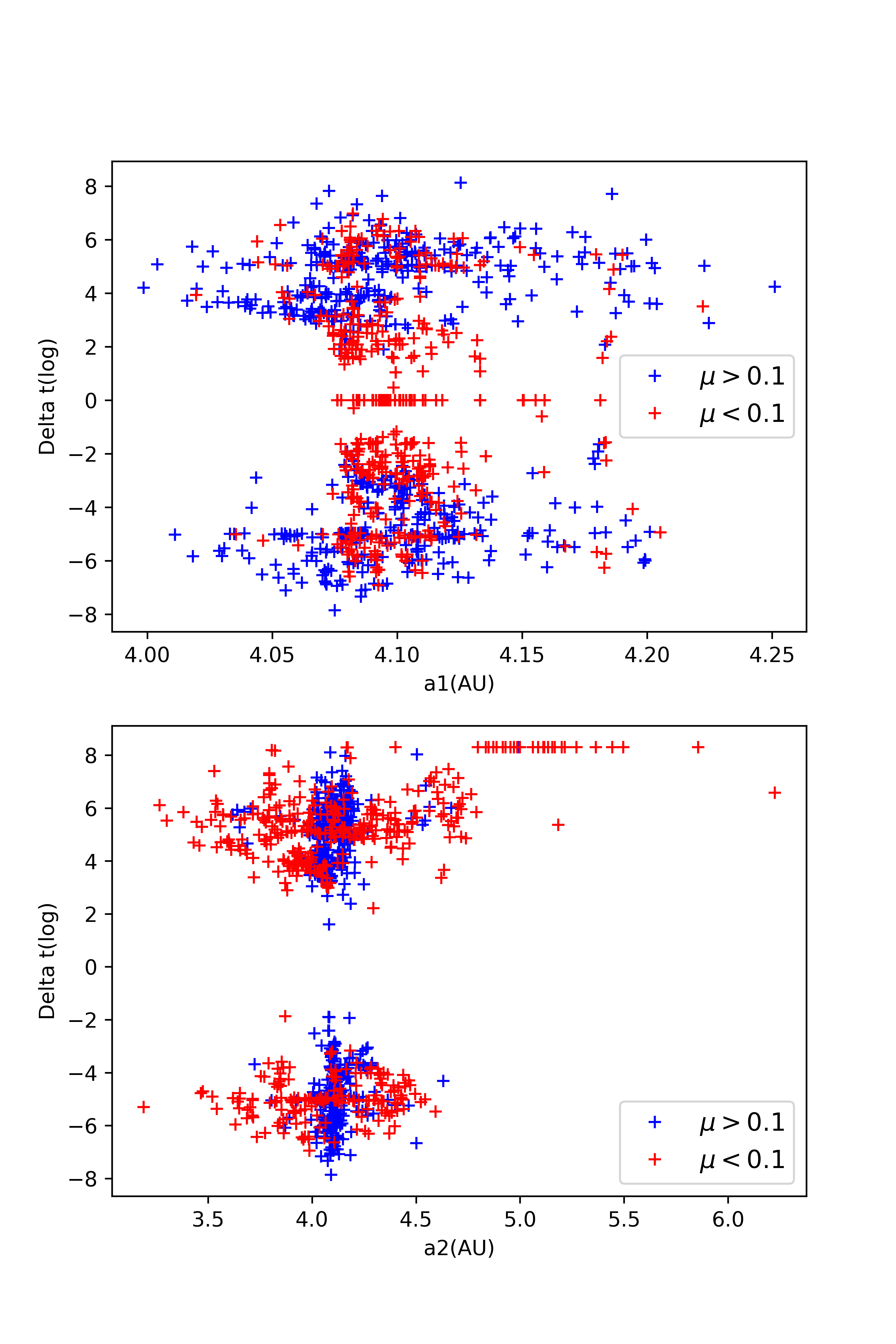}
    \caption{Change in the {accretion} time as a function of the orbital semimajor axes after binary dissociation for Ast1(upper) and Ast2(lower). Note:\ the scales in the x-axis are different in the two panels.} 
    \label{fig6}
\end{figure}

\begin{figure*}[]
    \centering
    \includegraphics[scale=0.7]{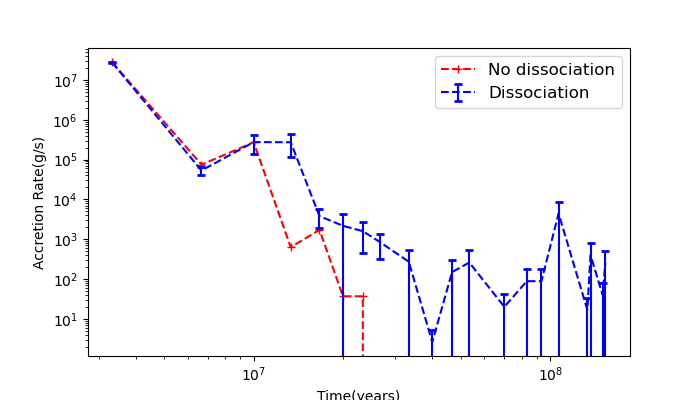}
    \caption{
    Effect of dissociation on the accretion rate. It shows the accretion rate as a function of time, 
    {error bar obtained assuming Poisson statistics.}
    The red dashed line is the case without dissociation and the blue dashed line is the case where dissociation is considered. }
    \label{fig:ac}
\end{figure*}

\section{Conclusions and Discussion
} \label{sec:Con}
In this work, we carry out several sets of simulations to investigate how asteroid binaries dissociate around a central WD and how this might affect the metal pollution of the WD. Our main conclusions are outlined below:

• Binary asteroids may dissociate when getting very close to a white dwarf and the two components end up on two different orbits.

• We derived an analytical expression for the semi-major axes of orbits of the components after dissociation and we show that it depends on the binary asteroids' total mass, the mass ratio, and the binary relative orbit. These are verified by our numerical simulations.

• The dissociation distance depends on the ratio between the binary total mass and WD mass and the binary relative orbit. Our numerical simulations show that the dissociation fraction at k=0.75 times the binary hill radius is about 0.5 and drops to 0 when k=1.5.

• The dissociation may change the time a component accretes onto the WD. Our simulations show that the change could be in either direction, a prolongation slightly favored, which is typically of the order of $10^5\sim10^6$ yr.

• The change in the accretion time is most significant when Ast2 is much smaller than Ast1, where the accretion of the former can be delayed by hundreds of millions of years. In this case, the change in the orbit of Ast2 is also the largest.

• We compared the WD accretion rate with and without taking into consideration the binary dissociation, finding that the overall effect of the asteroids' binarity is statistically insignificant: the rates in the two situations are in fair agreement with each other.

{In this work, we restrict our study to the effect that a binarity asteroid dissociation  may have on the WD accretion for asteroids from the inner main belt in the Solar System. In such a scenario, the asteroids' orbits are excited by the secular resonance, where the eccentricity increases and the semimajor axis is largely unaffected. As a consequence, those objects would experience scattering with the WD before Jupiter (if  at all possible). On the contrary, if the polluter binaries come from the other small body population, or the mechanism for sending them towards the WD is different, scattering with planets may be inevitable. \citet{10.1093/mnras/stad382} showed that scattering with the planets may cause the binaries' dissociation and ejection, the rate depending on the planets' mass and the system configuration. If the binaries are indeed dissociated by the planets before being flung close to the WD, our discussion becomes irrelevant. However, in neither case do we find that the influence on the WD accretion rate is significant.}

In our grid simulations in Sect. \ref{sec:ms2}, we see that those with the larger total semi-major axis are more affected  by binary dissociation. Such a phenomenon may have some implications {on} the accretion of objects from large distances. For a gaseous disc that extends much beyond its Roche limit, capturing a planetesimal from exo-Kuiper or exo-Oort cloud is more probable than tidal disruption at the Roche limit \citep{2019MNRAS.489..168G}. 
If a pair of binary planetesimals {(from the exo-Oort cloud)} are dissociated when they get close to WD, the semi-major axes may change by more than $10^4 AU$, therefore quickly moving the objects {apastron} down.  Such chang{es} may also have an effect on Alfvén-wave drag\citep{2021ApJ...915...91Z} {that act on} the fragment {from} large distance: such fragments have a long timescale of accretion and binaries dissociation may shorten the timescale.

Our results with regard to the {accretion} simulation shows that in some cases, the component from a dissociated binary may collide with the WD earlier than if the binary is taken as a single body by a time of the order of {$10^5 \sim 10^6 yrs$}. Simulations show that young ($<10^7yrs$) WDs have very low accretion rates \citep{2018MNRAS.476.3939M}. {Here, the deferment caused by the asteroid's binarity is shorter than the cooling age of these WDs so the effect is minor.}

We have omitted the evolution of the binary asteroids during the host star's main sequence and giant branch phase. Binaries may be destroyed by small impacts {with} other asteroids or scattering with planets. 
Tighter binaries have relatively good chances of survival. They may represent a trace of lucky survivors of a much larger population of the original disk binaries or they formed at $30AU$-$40AU$ and dodged the impact- and encounter-related perturbations that we studied here. Nearly all known satellites of the largest KBOs would have survived during the dynamical implantation of these bodies in the Kuiper belt \citep{2019Icar..331...49N}. During the post-main sequence \citep{2014MNRAS.445.2794V}, giant star radiation will destroy nearly all bodies with radii in the range 100m-10km that survive their parent star's main-sequence lifetime within a distance of about $7AU$, and binaries outer than $7AU$ are likely to survive. 
Although our model which is based on the Solar System is not consistent with such a range of survival, it is possible that binaries take a significant proportion of the small body population in extrasolar WD systems.

Another point that ought to be discussed is the mechanism responsible for the excitation of the eccentricity of asteroids. In our model, we only considered the $\nu_6$ resonance as the driver for the asteroid's high eccentricity, however, there are other mechanisms that can also cause a very high eccentricity, such as the {Kozai-Lidov effect} \citep{2016MNRAS.462L..84H,2016ARA&A..54..441N,2017ApJ...844L..16S} 
or mean motion resonance \citep{2012ApJ...747..148D}. In {the former} case, the asteroid's orbital {periastron} distance shrinks but the semimajor axis does not change much. Thus, if the asteroid is indeed a binary, our mechanism would operate equally well in the Kozai-Lidov case. However, if the asteroid is scattered close to the WD by a planet \citep{2014MNRAS.439.2442F,10.1093/mnras/stad382}, the scattering may already break the asteroid in the case when it is indeed  a binary. This would {therefore nullify} our mechanism.

\begin{acknowledgements}
We thank the anonymous referee for insightful comments that help improve the quality of this work. ZJ and ZZ was supported by the National Natural Science Foundation of China under Grants Nos. 11633001, 11920101003 and 12021003, the Strategic Priority Research Program of the Chinese Academy of Sciences, Grant No. XDB23000000 and the Interdiscipline Research Funds of Beijing Normal University. DL is supported by the National Natural Science Foundation of China (grants 12103007 and 12073019) and the Fundamental Research Funds for the Central Universities (grant 2021NTST08).
\end{acknowledgements}

\begin{appendix}
\section{Derivation} \label{ap1}
In Sect. \ref{sec:ms2}, We only provide the derivation without considering the effect of $\mu$, so here we give a detailed derivation about the relationship of $\mu$ and dissociation. 
If we use the specific expression of $v_{1}$ and $v_{2}$ such that:
\begin{equation}
v_{1} = \sqrt{\frac{G\left({m}_{2}^{2}\right)}{a_{rel}(m_1+m_2)}} = \frac{m_{2}}{m_1+m_2} v_{rel}
,\end{equation}
\begin{equation}
v_{2} = \sqrt{\frac{G\left({m}_{1}^{2}\right)}{a_{rel}(m_1+m_2)}} = \frac{m_{1}}{m_1+m_2} v_{rel}
.\end{equation}
Then the energy expression for Ast1 and Ast2 is:
\begin{equation}
\frac{\left(v_{tot} \pm \frac{m_{2}}{m_1+m_2} v_{rel}\right)^2}{2}-\frac{G M}{r_d \pm \frac{m_2}{m_1+m_2} \cdot a_{rel}}=-\frac{G M}{2 a_1}
,\end{equation}
\begin{equation}
\frac{\left(v_{tot} \pm \frac{m_{1}}{m_1+m_2} v_{rel}\right)^2}{2}-\frac{G M}{r_d \pm \frac{m_1}{m_1+m_2} \cdot a_{rel}}=-\frac{G M}{2 a_2}
.\end{equation}
The expression in Equation \eqref{eq m1} is changed as follows:
\begin{equation}
\pm\left(\frac{m_2}{m_1+m_2}\right)\left(\frac{m_1+m_2}{M}\right)^{2 / 3} \frac{1}{a_{rel}} \sim-\frac{1}{2}\left(1 / a_{tot}-1 / a_1\right)
,\end{equation}
\begin{equation}
\pm\left(\frac{m_1}{m_1+m_2}\right)\left(\frac{m_1+m_2}{M}\right)^{2 / 3} \frac{1}{a_{rel}} \sim-\frac{1}{2}\left(1 / a_{tot}-1 / a_2\right)
.\end{equation}
Or we can express this with $\mu$:
\begin{equation}
\pm\left(\frac{\mu}{1+\mu}\right)\left(\frac{m_1+m_2}{M}\right)^{2 / 3} \frac{1}{a_{rel}} \sim-\frac{1}{2}\left(1 / a_{tot}-1 / a_1\right)
\label{eq mu a1 app}
,\end{equation}
\begin{equation}
\pm\left(\frac{1}{1+\mu}\right)\left(\frac{m_1+m_2}{M}\right)^{2 / 3} \frac{1}{a_{rel}} \sim-\frac{1}{2}\left(1 / a_{tot}-1 / a_2\right)
\label{eq mu a2 app}
.\end{equation}
These expressions can explain the effect brought on by the mass ratio.
\end{appendix}

\end{document}